# On the origin of the crescent-shaped distributions observed by MMS at the magnetopause

G. Lapenta[1], J. Berchem[2], M. Zhou[2], R. J. Walker[3], M. El-Alaoui[2], M. L. Goldstein[4], W. R. Paterson[5], B. L. Giles[5], C. J. Pollock[5], C. T. Russell[3], R. J. Strangeway[3], R. E. Ergun[6], Y. V. Khotyaintsev[7], R. B. Torbert[8], J. L. Burch[9]

Correspondence to: Giovanni Lapenta giovanni.lapenta@kuleuven.be

1. Departement Wiskunde, KU Leuven, University of Leuven, Leuven, Belgium

2. Department of Physics and Astronomy, University of California, Los Angeles, CA, USA

3. Department of Earth, Planetary and Space Sciences, University of California, Los Angeles, CA, USA

4. Space Science Institute, Boulder, CO, USA

5. NASA, Goddard Space Flight Center, Greenbelt, Maryland, USA

6. University of Colorado LASP, Boulder, Colorado, USA

7. Swedish Institute of Space Physics, Uppsala, Sweden

8. University of New Hampshire, Durham, New Hampshire, USA

9. Southwest Research Institute, San Antonio TX, USA

Key Points (140 characters or less):
- Electron and ion velocity distributions have crescent shapes in opposite directions

- The magnetopause electric field is not a determining factor in forming the crescent distributions but it alters the extent of the crescent
- Crescent distributions are caused by the meandering particles in the magnetic field reversal of thin current sheets.

Index Terms and Keywords:

2724 Magnetopause and boundary layers

7807 Charged particle motion and acceleration

2723 Magnetic reconnection

7833 Mathematical and numerical techniques

**Abstract**

MMS observations recently confirmed that crescent-shaped electron velocity distributions in the plane perpendicular to the magnetic field occur in the electron diffusion region near reconnection sites at Earth's magnetopause. In this paper, we reexamine the origin of the crescent-shaped distributions in the light of our new finding that ions and electrons are drifting in opposite directions when displayed in magnetopause boundary-normal coordinates. Therefore, $\boldsymbol{E} \times \boldsymbol{B}$ drifts cannot cause the crescent shapes. We performed a high-resolution multi-scale simulation capturing sub-electron skin depth scales. The results suggest that the crescent-shaped distributions are caused by meandering orbits without necessarily requiring any additional processes found at the magnetopause such as the highly asymmetric magnetopause ambipolar electric field. We use an adiabatic Hamiltonian model of particle motion to confirm that conservation of canonical momentum in the presence of magnetic field gradients causes the formation of crescent shapes without invoking asymmetries or the presence of an $\boldsymbol{E} \times \boldsymbol{B}$ drift. An important consequence of this finding is that we expect crescent-shaped distributions also to be observed in the magnetotail, a prediction that MMS will soon be able to test.

## I. Introduction

Recently, *Burch et al.* [2016] reported a crossing of an electron diffusion region (EDR) by the Magnetospheric Multiscale (MMS) spacecraft. They used several diagnostics to confirm their conclusion. One feature used to identify the EDR was the observation of crescent-shaped electron distributions in the velocity plane perpendicular to the local magnetic field. Such crescent-shaped distributions were theoretically predicted before being observed by MMS [*Hesse et al.*, 2014] and are expected to occur in the region between the X-point and the flow stagnation point [*Bessho et al.*, 2016]. At the dayside magnetopause, the stagnation point is separated from the X-point, lying on the magnetospheric side of it, because of the asymmetric configuration of the reconnection topology [e.g., *Cassak and Shay*, 2007; *Pritchett*, 2008].

Crescent-shaped distributions have been associated with the meandering orbits of particles in a region of small magnetic field near a field reversal [*Lee et al.*, 2004]. Previous analyses investigated similar physical processes for ions, in what was in that case called "lima-bean"-shaped ion distributions ("lima-bean" shaped is just another way of referring to a crescent) [*DeCoster, Frank, 1979; Ashour-Abdalla et al, 1993; Frank et al, 1994*]. These ion distributions have been confirmed also by MMS [*Burch and Phan, 2016; Khotyaintsev et al., 2016; Phan et al., 2016*].

A number of papers have provided a theoretical explanation of the crescent-shaped distributions. As mentioned above, crescents were predicted by *Hesse et al.* [*2014*] based on PIC simulations, before even being observed. *Hesse et al* [*2014*] suggested that crescents in their PIC simulations were linked with meandering particle orbits. This conclusion was confirmed by the more detailed analysis of *Bessho et al.* [2016] who traced test particles using the fields from a full PIC simulation. They proposed a model for the crescents based on the assumption of a magnetic field reversing around the x-point and an asymmetric electric field zero sunward of the neutral point and linearly growing from the neutral point towards the stagnation point. *Shay et al.* [2016] proposed a model of crescents based on the presence of electric and magnetic fields, in this case with both reversing around the x-point. The crescent-shaped distributions in their model are characterized by a free parameter, the $\boldsymbol{E} \times \boldsymbol{B}$ drift velocity. These models focused attention on the presence of both an electric field and a magnetic field reversal.

In this paper, we first reexamine the origin of the crescent-shaped distributions observed by *Burch et al.* [2016]. By displaying the MMS data in the magnetopause's boundary normal (LMN) coordinate system, we show that ions and electrons are drifting in opposite directions. Since the motion of the ions and of the electrons are opposite, this rules out the concept that both species are $\boldsymbol{E} \times \boldsymbol{B}$ drifting, and in fact closer inspection of the MMS data shows that neither are drifting at the $\boldsymbol{E} \times \boldsymbol{B}$ drift velocity. Next, we present the results of multiscale simulations that include large features of a realistic state of the magnetopause while resolving the electron physics at sub-electron skin depth scales. The results reproduce the MMS data: ion and electron velocity distributions show crescent patterns in opposite directions, with drifts that are consistent with the local average particle speed.

Finally we provide a general theory of the particle motion based on the adiabatic Hamiltonian method [*Grad*, 1961; *Schmidt*, 2012]. Starting with a generalization of the models of *Bessho et al* [2016] and *Shay et al* [2016], we allow both fields to be a general linear function, allowing for the electric and magnetic field to better represent the actual conditions at the magnetopause where the electric field is neither constant nor simply reversing sign at the x-point and the magnetic field is asymmetric. This new model allows both for symmetric and asymmetric conditions. We use that approach to show that while the electric field is important in determining the span of the particle orbit in the Sun-Earth direction, it is not necessary for determining the presence of meandering orbits that result in crescent-shaped velocity distributions in the perpendicular plane. The perpendicular distribution is determined by the magnetic field vector potential via the law of conservation of canonical momentum in the dawn-dusk direction. The theory produces electron and ion distributions drifting in opposite directions, as observed in the MMS data without invoking ***E×B*** drift. Furthermore, no asymmetry is necessary in this model for determining the presence of crescent-shaped distributions.

The outcome of our analysis provides a straightforward explanation for the origin of the crescent-shaped distributions: they are enabled by the presence of a magnetic field reversal occurring on a length scale of the order of the particle's gyroradius. Because the EDR is a region of intense electron current around a magnetic null, this region is host to meandering orbits that result in crescent-shaped electron velocity distributions. This explanation removes the role of any peculiarities of the magnetopause and suggests that crescent-shaped distributions should be present also in the magnetotail. We report a result of a magnetotail simulation that confirms that hypothesis. Such a prediction can be readily tested when MMS explores the magnetotail.

## II. Observations of Crescent Distributions

The data used in our study were acquired when MMS was in burst mode. The Fluxgate Magnetometer (FGM) provides three-dimensional magnetic fields with cadence of 128 samples/s in burst mode [*Russell et al.*, 2014; *Torbert et al.*, 2014]. The Fast Plasma Instrument (FPI) provides 3-D electron distributions with a time resolution of 30 ms and ion distributions with a time resolution of 150 ms. The energy ranges of FPI are from 10 eV to 30 keV for both electrons and ions. Plasma moments (density, velocity, temperature, etc.) integrated by using the full distributions are also provided with the same cadence [*Pollock et al.*, 2016].

Figure 1 shows 2s (13:07:01 to 13:07:03) of data from an MMS2 crossing of the dayside magnetopause on October 16, 2015 when an EDR crossing was identified [*Burch et al., 2016*]. We have marked the time of the EDR crossing with a dashed line labeled A. All quantities are displayed in the LMN boundary-normal coordinates determined by *Burch et al.*, [2016]. These directions were determined from a minimum variance analysis of the magnetic field data between 13:05:40 and 13:06:09 UT. The EDR can be identified by the reversal of the electron flow speed component along the L and N component. There is an offset in the reversal of the ion drift speed due to the location of the x-point south the subsolar point (at $Z_{GSM}$=-4.7$R_E$). The EDR crossing was also characterized by strong electron heating and strong current dissipation, as indicated by $\mathbf{J} \cdot (\mathbf{E} + \mathbf{V}_e \times \mathbf{B})$, by the presence of a reconnection electric field and of regions of violation of the electron frozen in condition [*Burch et al.*, 2016]. Additional supporting evidence was found in the observation of clear crescent signatures in the electron velocity distribution as predicted in previous simulation studies [*Hesse et al.*, 2014].

Figure 2 shows the ion distribution measured over 150 ms and the electron distribution measured over 30 ms at the time marked A in Figure 1, which corresponds to the time when the spacecraft were in the vicinity of the EDR, as indicated by the three components of the magnetic field being nearly zero [*Burch et al., 2016*]. The distributions are shown in the $V_M$-$V_N$ velocity plane while the MMS software uses axes in the perpendicular velocity plane defined by $V_{perp1} = (\mathbf{b} \times \mathbf{v}_s) \times \mathbf{b}$ and $V_{perp2} = \mathbf{v}_s \times \mathbf{b}$ where $\mathbf{b}$ and $\mathbf{v}_s$ are the unit vectors along the magnetic field and the species velocity and therefore are different for ions and electrons. The LMN system of coordinates is independent of the species and thus allows us to compare electron and ion data in the same coordinate system. Figure 2 shows that the electrons form a crescent-shaped distribution in the normal $V_M$-$V_N$ plane and that the ions drift in the direction opposite to the electrons. The key observation is that electrons and ions drift in opposite directions. Since direction and magnitude of an $\mathbf{E} \times \mathbf{B}$ drift is independent of the species, it cannot explain the different directions of the electron and ion crescents.

While ions are not expected to be magnetized, a close inspection of Figure 1 reveals that the ions are not following the $\mathbf{E} \times \mathbf{B}$ drift and neither are the electrons. At time A, the $V_{eM}$ velocity in Figure 1 shows a positive peak above 1000 km/s consistent with the electron distribution of Figure 2. At this time, Figure 1 bottom panel shows that the $\boldsymbol{E} \times \boldsymbol{B}$ drift in the M direction is much smaller than $V_{eM}$. For this same period, the N-component of the drift speed is completely different and has opposite sign from the $\boldsymbol{E} \times \boldsymbol{B}$ drift speed. During the whole interval shown the ion velocity does not follow the $\boldsymbol{E} \times \boldsymbol{B}$ drift as the ions remain unmagnetized.

The actual measured speeds are reported in Table 1. The ion and electron sampling rates are different so the interval of measurement for electrons and ions is not identical, although it overlaps. The bulk speeds measured are different from the $\boldsymbol{E} \times \boldsymbol{B}$ drift. The diamagnetic drift is measured for each species as

$$\boldsymbol{V}_{d,s} = -\frac{\nabla \cdot \mathbb{P}_s \times \boldsymbol{B}}{q n_s B^2}$$

by using the full pressure tensor. As can be observed, the bulk speeds for both species differ greatly from either the diamagnetic or the $\boldsymbol{E} \times \boldsymbol{B}$ drift. The diamagnetic drifts are especially off. This finding might at first appear puzzling and it is possible that averaging methods might lead to more realistic numbers. But comparing with simulations reported below, the same large discrepancy is observed also in the simulations. The real reason is that in the vicinity of the stagnation point the divergence of the pressure tensor is especially large, almost becoming a singularity. This same behavior is derived both from MMS data and from the simulation results.

### III. Magnetopause crossing in multiscale simulations

We have applied our recently developed multiscale approach to capture the microphysics while accurately taking into account the macroscopic state of the magnetosphere [*Ashour-Abdalla et al.*, 2015; *Lapenta et al.*, 2016]. The approach consists of isolating a subset of a magnetospheric state predicted by a global MHD simulation and using it as initial state for a high-resolution particle-in-cell (PIC) simulation that fully resolves the kinetic behavior of both ions and electrons within that small domain. This is not a coupled simulation but rather a one-way communication that occurs at only one selected time. From that time the PIC run continues completely independently, with full physical description of its domain. The simulation results are displayed in the local LMN coordinate system at the subsolar magnetopause. The transformation from LMN to GSE coordinates is straightforward: L= z, M=-y. and N=x. Using that local system of coordinates allows us to compare simulation results at the subsolar point with MMS measurements in the afternoon sector. Hence, all simulation results reported in the reminder of the paper are displayed in LMN coordinates where lengths are normalized to the reference ion skin depth in the magnetosheath $d_{i,sh}$ and velocities to the speed of light.

The initial state of the PIC code was taken from the results of a 3D global MHD simulation [e.g., *Raeder et al.*, 1995; *Berchem et al.*, 1995]. The MHD code solves the one-fluid MHD equations and includes the resistive form of Ohm's law [*Raeder et al.*, 1996]. For this study, we used a rectangular but non-uniform grid over a large simulation domain (200×80×80 $R_E$) with a resolution of about 0.025 $R_E$ (160 km) in the subsolar region. To simplify the configuration of the interaction of the solar wind with the magnetosphere, we neglected the tilt of the Earth's magnetic dipole (hence the *L*-axis of the simulation system coincides with the *z*-axis of the geocentric GSE and GSM coordinate systems) and used a uniform Pedersen conductance of 5 S and zero for the Hall conductance.

Values of the IMF ($B_N = B_M = 0$; $B_L = -8$nT) and solar wind plasma velocity ($V_N = -650$ km/s, $V_L = V_M = 0$), density ($n_{swd} = 4$ cm$^{-3}$) and solar wind temperature ($T_{swd} \approx 600$ eV) were imposed at the righthand boundary of the simulation system. The main magnetopause current layer obtained in the MHD simulation was about 500 km thick in the subsolar region. Approximate values of the fields and ion plasma parameters upstream and downstream at the subsolar magnetopause were: $d_i = 70/600$ km, $\rho_i = 120/50$ km, $T_i = 1.3/2.8$ keV, $n_i = 10/0.04$ cm$^{-3}$, $V_N = -100/50$ km/s, and $B_L = -30/90$ nT; hence $\omega_{pi} = 650/50$ Hz, and $\omega_{ci} = 0.45/1.4$ Hz where the numbers given are the magnetosheath values over the magnetosphere values. The results reported below do not depend critically on the details of the parameters chosen for the global MHD simulation. We conducted several runs with different parameters and the conclusions below are independent of the details of the choices made.

Two sets of 2D kinetic simulations were carried out by using the iPic3D code [*Markidis et al.*, 2016]. While the results of the MHD simulations in the noon-meridian were used to set the initial conditions for both PIC simulations, we used two different levels of resolution. Run A had a domain of 185 × 370 $d_{i,sh}$ and Run B had a zoomed-in domain of 26 × 52 $d_{i,sh}$ (SH refers to the magnetosheath and SP to the magnetosphere). Both domains were centered on the magnetopause subsolar region and had 1600 ×3200 cells. The simulations were run using 3200 processors of NASA's Pleiades and Discover supercomputers. We employed open and non-reflecting boundary conditions on all boundaries except the lefthand boundary where we imposed a constant plasma inflow using the speed from the selected MHD state. The electron skin depth and particle motion were well resolved in the entire domain of both simulation runs. The grid spacing (equal along L and N) ranges from $\Delta s=0.6\ d_{e,sh} = 0.08\ d_{e,sp}$ in Run A to $\Delta s=0.15\ d_{e,sh} = 0.02\ d_{e,sp}$ in Run B. The time step is $\omega_{ce,sh}\ \Delta t= 0.02$, or equivalently $\omega_{ce,sp}\ \Delta t= 0.08$. Both runs used the same temporal resolution. While the spatial resolution was uniform, the variation of the effective electron skin depth and gyrofrequency at the dayside magnetopause are such that the spatial resolution was better in the magnetosphere because of the low density and the temporal resolution was better in the magnetosheath because of the low magnetic field. The local resolution of the region of low density and nearly vanishing magnetic field which characterizes the EDR was excellent in the B run: $\Delta s=0.2\ d_{e,EDR}$ and $\omega_{e,meandering}\ \Delta t= 0.02$ (using the meandering bouncing frequency rather than the gyrofrequency) and adequate in the A run.

To have sufficient electron resolution with the available resources, we employed a mass ratio $m_i/m_e=25$, which has been used in earlier PIC simulations of crescent distributions at the magnetopause [e.g., *Bessho et al.*, 2016]. Previous work has shown that a reduced mass ratio has a quantitative but not qualitative effect: the physics observed is the same but the precise numbers differ for a physical mass ratio [e.g., *Ricci et al.*, 2002; *Lapenta et al.*, 2010]. This finding is supported by the reconnection mass-ratio scaling laws [*Shay et al.*, 1999; *Cassak and Shay*, 2007].

Panels (a), (b) and (c) of Figure 3 show values of $Q_e$, the parameter that is used to quantify the electron agyrotropy (i.e., non-gyrotropy) for Runs A and B. They were obtained by using *Swisdak*'s definition [*Swisdak*, 2016]. As we discuss below, the values of $Q_e$ computed by using the two other definitions of agyrotropy proposed by *Scudder and Daughton* [2008] and *Aunai et al.* [2013] do not differ significantly from these results. Panels (a) and (b) show results from the lower resolution Run A (the whole domain is shown in panel (a) with a close-up of the subsolar magnetopause in panel (b)) while panel (c) shows results taken from the higher resolution Run B. The EDR can be identified by the very large values of $Q_e$ that prevail earthward of the X-point. The X-point can be identified as the converging point of the magnetic field lines (black traces) at N =11.75 $d_i$.

Figure 4 reports different components of the ion and electron velocities, the electric field, and the current. The stagnation points observed in Panels a and b are located where the normal speed ($V_N$) vanishes. The electrons and ions do not reverse their velocity in a smooth univocal way, but rather significant structure is present. While the stagnation point is precisely defined for the electrons at N/$d_i$=11.24, it occurs in a more diffuse area for the ions. A strong in-plane ambipolar electric field (Figure 4-c) and the out-of-plane current seen in panel (Figure 4-f) are caused by the electron-ion charge separation at the magnetopause. Although both the electric field and the current are located earthward of the EDR, they are not directly related to the occurrence of reconnection. Ambipolar electric fields develop at the magnetopause without requiring any reconnection [e.g., *Berchem and Okuda*, 1990] and electrons and ions deflected in opposite directions by the magnetospheric field create a strong current [*Chapman and Ferraro*, 1931]. The ripple in the electron motion (Figure 4-a) results from an instability created by the ambipolar electric field. Panels (d) and (e) in Figure 4 display the ion and electron perpendicular drifts, respectively. The ions penetrate deeper into the magnetosphere by virtue of their larger gyroradii and, as expected, drift in the direction opposite to that of the electrons. The strong in-plane field causes a clear ***E***×***B*** drift in the M-direction for the electrons (Figure 4-e) but ***E***×***B*** is not dominant for the ions (Figure 4-d).

Panel (a) in Figure 5 displays a series of stacked high-resolution profiles of $V_{eN}$ (blue), $V_{iN}$ (red), the N component of the ***E***×***B*** drift (violet) and of the electron diamagnetic drift (yellow), $E_N$ (green) and $B_L$ (black) in the top subpanel.

The electron bulk speed (blue) differs radically from the ***E***×***B*** drift (violet) and the diamagnetic drift (yellow). The latter varies widely in the vicinity of the diffusion region, a fact we already noted in the analysis of the MMS data. While the explanation of the crescent in terms of diamagnetic drifts is as tempting as that on the ***E***×***B*** drift, neither appears to be tenable based on panel (a) of Figure 5.

The values of the N and M components of the non-ideal terms in the Ohm's law in the middle subpanel of Figure 5a indicate where the frozen-in condition is violated. The bottom subpanel of Figure 5a contains the agyrotropy calculated by using the 3 methods mentioned above. These cuts through the EDR are taken along the N-direction at $L/d_i$ =25.68 for Run B. In the top subpanel, the reversal of the northward component of the magnetic field ($B_z$) at N =11.75 $d_i$, and the region of low earthward velocities ($V_{eN}$) near N =11.2 $d_i$ mark the location of the X-point and stagnation point, respectively. In this region, the electron and ion drift speeds in the N direction are completely different from the $\boldsymbol{E} \times \boldsymbol{B}$ drift.

The electric field $E_N$ exhibits a complex two-peak structure corresponding to the X-point and stagnation point with some enhancement in the region of the Chapman-Ferraro current. In the middle subpanel of Figure 5a the violation of the frozen-in condition indicates that the out-of-plane component of $\boldsymbol{E} + \boldsymbol{V}_e \times \boldsymbol{B}$ (denoted $OHM_M$ in red) is enhanced in the region between the X-point and the stagnation point. It can be used to identify precisely the EDR [*Pritchett and Mozer*, 2009; *Hesse et al.*, 2014]. However, it increases again in the region of the Chapman-Ferraro current (L~10.5), a process also linked with breaking the frozen-in-flux condition but not related to reconnection.

Panel (b) in Figure 5 shows electron velocity distributions in the plane ($V_N$, $V_M$) perpendicular to the vertical magnetic field $B_L$ taken at different N locations for Run B. The distributions were obtained by using the particles in a square box of size $0.15 d_{i,sh}$ (corresponding to a local electron skin depth). Crescent-shaped distributions similar to those observed by MMS are present in the entire EDR in the region spanning the X-point and the stagnation point (see also *Bessho et al.* [2016]).

Panel (c) shows the results for ions by using the same method we used for electrons. Moving from the magnetosheath to the magnetosphere we pass through the X-point where we see two beams traveling from the magnetosheath (left in Fig. 5) and the magnetospheric (right in Figure 5) sides of the X-point. The magnetospheric beam is weaker because of the lower density in the magnetosphere. The beams merge and a crescent shape forms on the magnetospheric side. The location of the fully developed ion crescent is further into the magnetospheric side than the electron crescent.

After crossing the EDR, MMS entered the reconnection outflow region [*Burch et al.*, 2016], hence the spacecraft did not penetrate far enough into the magnetosphere to observe a fully developed ion crescent. Nevertheless, the simulation results are consistent with the fact that the MMS data show that the electrons forming the crescent distribution are drifting in the direction opposite to that of the ions. The crescent distributions form where there are sufficiently steep magnetic field gradients such that the particles can follow meandering orbits. The steep field is associated with a strong current caused by the deflection of the incoming magnetosheath particles produced by the strong ambipolar electric field [*Chapman-Ferraro*, 1931; *Willis*, 1971].

### IV.  Single particle theory of crescents

Recently, *Bessho et al.* [2016] provided a description of the formation of crescent distributions by solving Newton's equations for single particle motion in presence of both a linear magnetic field reversal at the X-point and an in-plane electric field directed along x that is asymmetric: zero until the X-point and linearly growing from it toward the stagnation point. They reproduced the crescent distributions correctly, but the role of the ***E×B*** drift was dominant. *Shay et al.* [2016] also suggested that ***E×B*** drift could result in the formation of crescent shaped distributions.

However, as we have shown above, our simulation results suggest that the electric field does not play a crucial role and the electron crescent-shaped distributions simply result from the presence of meandering orbits caused by the reversal of the magnetic field in a narrow electron-scale region. The outstanding question then is to determine whether single particle theory can explain crescent-shaped distributions when the electric field is absent or non-uniform. To answer this question, we use adiabatic Hamiltonian theory of single particle motion [*Grad*, 1961; *Northrup*, 1963; *Sonnerup*, 1971; *Schmidt*, 2012] since it lends itself better to describe general field geometries and interpret particle trajectories than solving Newton's equations.

Let's assume a model where the magnetic field **B** is directed along z (L in the results above) and varies only along x (along N above) with the field reversal occurring at the origin of the coordinate system x=0. A non-uniform electric field **E** is directed along x to represent the strong normal electric field formed by the ion pressure gradient across the magnetopause [*Willis,* 1971]. Figure 6a illustrates the configuration.  As in *Bessho et al.* [2016], the assumption is made of invariance along the *y*-direction (M direction above). The electric field along *y* is too small to be considered.

Under these assumptions, the canonical momenta in the y and z directions are invariant and the Hamiltonian for the motion of particles of any species can be written in a 1D-like form as

$$H_s = \frac{p_x^2}{2m_s} + \psi_s(x)$$

The canonical momentum is $\boldsymbol{p} = (m_s v_x, m_s v_y + q_s A_y, m_s v_z)$ where $A_y$ is the vector potential and $q_s$ is the charge of species $s$. The pseudo-potential is:

$$\psi_s(x) = \frac{(p_y - q_s A_y)^2}{2m_s} + \frac{p_z^2}{2m_s} + q_s \varphi$$

The magnetic field is described as $B_z = dA_y/dx$ and the electric field is given by $E_x = -d\varphi/dx$ where $\varphi$ is the electric potential.

The Hamiltonian formulation in terms of pseudo-potentials allows us to describe the particle trajectory as quasi-1D motion and to derive analytical expressions for complex fields via the choice of $A_y$ and $\varphi$. Assuming $\boldsymbol{E}$ and $\boldsymbol{B}$ to be linear in the vicinity of the EDR (see Figure 6a), the resulting quadratic potentials are:

$$A_y = B_0 x + \frac{B_1}{2} x^2$$

$$\varphi = -E_0 x - \frac{E_1}{2} x^2$$

By choice of the coordinate system, we have imposed $B_0 = 0$ because x=0 corresponds to the field reversal. But it is possible for a non-zero electric field to be present at the magnetic reversal, allowing a non-zero $E_0$. At large distances from the field reversal the linear dependence of the magnetic field is no longer true (for example a hyperbolic tangent could be used) but our attention is limited to the immediate proximity of the magnetic reversal.

Figure 6b shows that the coordinate x and its conjugate canonical momentum behave like the classic particles bouncing off the walls of a confining potential $\psi_s(x)$ where the canonical momentum $p_x$ vanishes. When the pseudo-potential has a single central well, the particles cross the central x=0 location and perform meandering orbits. When the pseudo-potential well has two minima, the low energy (i.e., the low $p_x$) particles perform gyro-orbits that are entirely confined inside the well [*Schmidt*, 2012]. The invariant canonical momentum in the y direction and the two potentials, $Ay$ and $\varphi$ determine the shape of the pseudo-potential well and determine the turning points. Note that the only term present to break the left-right symmetry in our model is $E_0$. Figure 6b shows a specific example where the particle trajectory is determined in the presence of an electric field at the magnetic field reversal. The non-zero $E_0$ leads to a left-right asymmetry, which results in turning points that are not symmetric.

For given fields, and therefore potentials $\varphi$ and $A_y$, the invariant canonical momentum in the y direction determines whether the potential is single welled or doubled welled. Given that, the energy in the x and z directions determine how high the particle is located with respect to the minima in the potential well. For one range of $p_y$ values, particles move in a single potential well and their orbits are meandering, while for another range they are line tied and confined to a single side.

The precise transition from one class to the other can be computed for a given electric and magnetic field by requiring that the minimum (i.e., solving $\frac{d\varphi}{dx} = 0$) be unique. For $E_0 = 0$, the potential well is symmetric, a symmetry that is not broken by a non-zero value of $E_1$. However, the value of $E_1$ modifies the limit value for $p_y$, hence, determining whether the orbits are meandering or line tied.

The pseudo-potential governs the motion in the x-direction while the canonical momentum $p_y$ determines the shape of the velocity distribution in the perpendicular plane (Vx-Vy). For this reason, the presence of **E** is not important and its details are immaterial in determining the shape of the distribution. Figure 6 panels d and f illustrate how the particle velocity evolves in the Vx-Vy plane. The canonical momentum $p_y$ is invariant and the velocity $V_Y$ varies in response to the changes of $A_y$ as the particle moves in x. The vector potential $A_y$ is positive definite. Considering the case of electrons, $A_y$ increases the velocity and the particle velocity behaves as illustrated in Figure 6e forming a crescent with extremes indicated by the arrows. The minima are given by $\frac{p_y}{m}$ and two maxima are given by $\frac{p_y + A_y(x^\pm)}{m}$. When the average electric field $E_0$ is zero, the two values of the turning points $x^\pm$ are equal and opposite and the crescent becomes a unique arc. The symmetry in Vx is a consequence of the particles traversing the x span of the pseudo-potential first in one direction and then the other, encountering the same value of $A_y$ but opposite values of Vx.

The particle trajectories cannot reach the full 360 degrees in the Vx-Vy plane because $mv_y = q_s A_y - p_y$ is limited in the range of values allowed. From the definition, the sign of $q_s A_y$ is opposite for electrons and ions, forming a crescent that is shifted by the invariant canonical momentum $p_y$.

Panels 6d and f show a collection of traces for different choices of the invariant $p_y$, for a given particle but different for each particle. Panel d is for ions and panel f is for electrons. The meandering electrons and ions are confined to crescents with opposite curvature for electrons and ions. On average, each crescent pattern corresponds to a drift path of the particle. Since this drift is caused by the variation of Ay resulting from the presence of a gradient in the magnetic field, the cause of the crescent shapes is directly related to the presence of a strong inhomogeneity in the magnetic field. Hence, the **E×B** drift is not relevant in the crescent formation process.

This analysis provides three critical predictions for particles in meandering orbits:

1) The electron and ion velocity distributions perpendicular to the magnetic field have crescent shapes that are oppositely positioned with respect to the $V_y$ direction (M direction in the observational data).
2) The presence of an electric field or of its asymmetry is not a determining factor in forming the crescent distributions but its presence alters quantitatively the extent of the crescent as it affects the turning points of the orbits in the quasipotential.
3) Electron crescent patterns are particularly well formed near the EDR because of the extreme thinning of the current sheet near the X-point.

## V. Discussion

The main conclusion reached above is that the crescent shape distributions observed by MMS at the magnetopause are determined by strong magnetic field gradients that produce meandering orbits resulting in crescent-shaped distributions in the velocity space. While these magnetic shears are typically accompanied strong asymmetric electric fields, our model shows that these electric fields are neither necessary nor sufficient for the presence of crescent-shaped distributions. Hence, we should expect crescent distributions to be observed whenever a steep reversal of the magnetic field is present, since by Ampere's law these are regions of strong currents. Ion crescents will be formed when the current scale associated with the magnetic field reversal (i.e., assuming Ampere's law to link curl **B** with **J**) is on the ion scale [*Lee et al.*, 2004] and electron crescents when the current sheet is on the electron scale, as in the case of the EDR. Since these conditions are not unique to the magnetopause EDR, we can expect to observe crescent distributions in other regions of the magnetosphere.

For example, we should expect to observe crescent distributions at both north and south latitudes away from the reconnection region. However, we should expect the effect to be less pronounced because the scales of the magnetic field variation and the current layer increases from the electron to the ion scales away from the EDR. Figure 7 shows three different locations at $L/d_i=28.9$ (panels a), just above the EDR, and at $L/d_i=32.2$ (panels b) well above it and at $L/d_i=19.2$ (panels c), symmetrically located in the southern hemisphere. Crescent distributions are present outside the EDR but with greatly reduced intensity, progressively fading away with distance from the EDR. The two symmetric locations above and below (panels b and c) are virtually identical, a consequence of the fact that in absence of a guide field, as in the present case, the system retains a strong north-south symmetry. This conclusion is reinforced by other simulation results [*Chen et al, 2016; Egedal et al, 2016*] that also report crescent-shaped distributions on the separatrices, away from the EDR proper.

A second example, and a more important one in light of the magnetotail phase of the MMS mission, is the fact that the results of our analysis do not depend on the peculiar asymmetry of the reconnection process at the magnetopause. As long as electron-scale field reversals are present, electron crescent-shaped distributions should be observed. Hence, we should expect crescent distributions to be present also for symmetric reconnection geometries. To prove this point we carried out a symmetric reconnection calculation. In this case, the one-sided magnetospheric electrostatic field is replaced by the aintisymmetric Hall electric field [*Lapenta et al., 2015; Goldman et al., 2016*]. We used the typical setup of the GEM challenge [*Birn et al.,* 2001; *Ricci et al.*, 2002]: an initial current Harris current sheet with an additional background and no guide field. For a more direct comparison with Run B, we used a mass ratio of 25 and a temperature ratio of 10. The other parameters of the Harris sheet were: electron thermal speed $v_{th,e}/c=0.074$ and Harris layer thickness $\delta/d_i=0.5$, with a background density equal to 0.38 of the peak Harris density (a value similar to that observed on the magnetospheric side of Run B). Figure 8 displays the electron distribution function observed at four locations along the N direction normal to the Harris sheet and positioned symmetrically around the EDR, located at N=0 in this simulation. The crescent shape is obviously present, but weaker than the ones obtained for the magnetopause simulations. While this simulation does not represent a realistic magnetotail configuration because of the magnetopause's range of temperatures and densities were used, it proves the point that the crescent distributions observed are due to meandering orbits and not by any peculiarity related to reconnection asymmetries at the magnetopause. This consideration makes us confident that crescent–shaped electron distributions will be observed when MMS crosses the magnetotail current sheet.

## VI.     Summary and Conclusions

In this study, we have carried out three different analyses relevant to the formation of the crescent shaped velocity distribution functions.

First, we displayed the electron and ion distributions of a recent MMS EDR crossing [*Burch et al.*, 2016] using the LMN boundary-normal coordinate system. The choice of such a coordinate system allowed us to identify the first critical finding of our paper: electron and ions drift in opposite directions. This implies that the drift associated with the formation of the crescents cannot be an ***E*×*B*** drift and, in fact, a close inspection of the data reveals neither species is simply following an ***E*×*B*** drift.

Second, we conducted multi-scale and highly resolved PIC simulations initialized by a realistic state of the magnetopause obtained from a global MHD simulation model. The PIC simulations used domains that were centered on the subsolar magnetopause and followed the evolution of reconnection with resolution capable of resolving scales of the order if not better than the electron skin depth. We found that the simulations reproduced the electron distributions observed by MMS, and that the two populations were drifting in opposite directions. In addition, the simulations indicated that these strong and opposite drifts were caused by the deflection of the solar wind particles by the magnetospheric field and that the crescents formed in the region where the Chapman-Ferraro current is located. The simulation results provided the second critical finding of our research: crescents are formed by meandering particles and can be formed only where there are strong gradients of the magnetic field (and therefore currents) such as near a magnetic field reversal. Crescents are more difficult to form in regions of magnetic field gradients without strong reversals, for example in cases of reconnection in presence of a guide field.

Finally, we used an adiabatic Hamiltonian formulation to analyze single particle motion and showed that the model confirms the critical role played by the presence of a strong localized inhomogeneity in the magnetic field. The electric field, or its asymmetry, is not a requirement for having meandering orbits or crescent-shaped distributions. However, such orbits and distributions are populated by energetic particles. The electric field can, however, play an important role in injecting particles in these higher energy orbits.

The final conclusion is that the crescents recently seen by the MMS mission [*Burch et al.*, 2016] are caused by the meandering particles in the magnetic field reversal between the magnetosheath and magnetospheric sides of the current layer. While asymmetries and ambipolar electric fields are present at the magnetopause they are not critical in determining the formation of crescent-shaped distributions. Hence, crescent-shape distributions can be expected for symmetric reconnection configurations. *Wilber et al.* [2004] identified crescent-shaped ion distributions measured by Cluster in the magnetotail but Cluster did not have the temporal and spatial resolution to observe electron-scale current layers. We expect that MMS will quickly resolve this point by observing crescent-shaped electron distributions when exploring reconnection regions in the magnetotail current sheet.

**Acknowledgements**

This work was supported by a Magnetospheric Multiscale Mission Interdisciplinary Scientist grant (NASA grant NNX08AO48G) at UCLA, a NASA Geospace grant (NNX12AD13G) and a NASA Heliospheric Grand Challenges grant (NNX14AI16G). One of the authors (GL) acknowledges partial support from the Belgian Space Policy IUAP grant CHARM, from KULeuven BOF and GOA grants and from the EC project DEEP-

## TABLE 1

|  | $V_L$ [km/s] | $V_M$ [km/s] | $V_N$ [km/s] |
|---|---|---|---|
| Electron bulk velocity | 564 | 597 | -296 |
| Electron $E \times B$ speed | 73 | 325 | 98 |
| Electron diamagnetic speed | -1149 | -7780 | -14896 |
| Ion bulk velocity | -149 | -83 | -23 |
| Ion ExB speed | 420 | 860 | 143 |
| Ion diamagnetic speed | 1649 | 7201 | -5418 |

**FIGURE CAPTIONS**

Figure 1: Magnetic field components (a), ion and electron density (b), ion velocity components (c), electron velocity components (d), electric field components (e) and $\boldsymbol{E} \times \boldsymbol{B}$ drift (f) from MMS2. All vectors are in the LMN boundary-normal coordinates defined by *Burch et al.* [2016]. N is the normal to the boundary and points away from the Earth, L corresponds to the main magnetic field component and is perpendicular to N and in the plane of reconnection (nearly along the magnetospheric magnetic-field direction), and M is normal to the L, N plane (predominantly dawnward). The time labeled as A is identified with the EDR crossing [*Burch et al.*, 2016].

Figure 2: Electron and ion distribution functions at time A identified in Figure 1 displayed in magnetopause LMN boundary-normal coordinates defined by *Burch et al.* [2016]. The velocity plane $V_N$-$V_M$ is normal to the main magnetic field component $B_L$.

Figure 3: Characterization of the EDR from two multiscale simulations at the time $\omega_{ci,SP} t = 37$ (corresponding to the cyclotron frequency in the magnetosphere, or about 5 seconds of real time). Panels a and b are from Run A and panel c is from Run B (see the text for the parameters used in the runs). The non-gyrotropy values are computed by using the *Swisdak* [2016] definition. In all panels the black lines are magnetic surfaces computed from the out of plane vector potential. Run A shows the overall picture in the complete Run A domain (a) and in a zoom-in around the magnetopause subsolar region (panel b). Panel c shows the non-gyrotropy in a small subset of the Run B domain to show a close up of the EDR.

Figure 4: False color representation of the central region around the EDR observed in Run B. Panels a and b show the normal speed $V_{iN}$ and $V_{eN}$ for ions and electrons, respectively. Panel c shows the normal electric field $E_N$. The normal direction points sunward. Panels d and e show the out of plane perpendicular component of the velocity for ions and electrons, respectively. Panel f shows the out of plane (direction M) component of the electron current $J_{eM}$. The X-point is identified by the convergence of the magnetic

surfaces shown in black while the stagnation point is identified by a zero value for the N-directed velocity: the stagnation point of electrons and ions is different; the ion panels (a, d) show a larger subdomain.

Figure 5: Panel a shows the traces of three physical quantities from run B which were taken along N in the vicinity of the EDR, and computed at the L position of the X-point ($L/d_i$ =25.68). On the top portion of panel a, the normal electric field, $E_N$, the normal velocity for electron (ions), $V_{eN}$ ($V_{iN}$), the normal $V_{ExB,N}$ drift, the electron diamagnetic drift $V_{de}$ and the magnetic field, $B_L$, are plotted. The middle part of panel a shows the violation of the electron frozen-in condition, $\boldsymbol{E} + \boldsymbol{V_e} \times \boldsymbol{B}$ in the N and M directions. The bottom portion shows the three measures of non-gyrotropy discussed in the text (*Aunai et al.*, [2013] in blue, *Swisdak*, [2016] in red and *Scudder and Daughton* [2008] in orange). Panels b and c show the electron and ion velocity distributions in the perpendicular plane, respectively. The distributions are measured in a small box of size $0.15d_{i,sh}$ around the position indicated above each distribution. The center of mass of the particles in the box is shown above each plot.

Figure 6: Adiabatic Hamiltonian model of the particle motion around the EDR. Panel a shows the idealized (linear) fields assumed in the model. Panel b shows one specific example of the pseudo-potential of the Hamiltonian model, the blue line is the vector potential $A_y$ and the black line is the pseudopotential ψ derived from it. The model is not sensitive to the values, but just for presentation we chose specifically $B_0$=0, $B_1$=1, $E_0$=-1, $E_1$=0. A particle is confined by the well and oscillates between $X^-$ and $X^+$ where the vector potential Ay takes the value indicated by the violet arrows. Panel c shows the x-$V_x$ phase space and panel e the perpendicular velocity plane Vx, Vy for a particle. The particle forms a crescent shaped figure in the velocity plane, the value of the canonical momentum $p_y$ and of the vector potential at the two end points ($X^-$ and $X^+$) define the crescent. Panels d and f show a number of velocity plane trajectories of selected particles forming a crescent. The ions (red, panel d) and electrons (blue, panel f) are oppositely directed.

Figure 7: Electron velocity distribution in the perpendicular plane measured as in Figure 5b, in a small box of size $0.15d_{i,sh}$ around the position indicated above each distribution. The distributions were determined at the same locations along the L-direction (indicated above each distribution) as in Figure 5 but at different north-south positions. Three sets of panels are shown. Panels in row a are taken around the $L/d_i= 28.92$, just north the EDR. Panels in row b and c are taken at two symmetric positions north and south of the EDR: $L/d_i= 32.17$ (north of the EDR) and $L/d_i= 19.17$ (south of the EDR).

Figure 8: Electron distribution function in four different locations above and below the X-point (located at N=0) of a simulation for a symmetric reconnection geometry, which used a plasma with the same beta and temperature as that found in the magnetospheric side of Run B to form a Harris equilibrium (see text for detail). The plots are displayed in the perpendicular plane $V_M$-$V_N$. Velocity is normalized by the speed of light. The crescent shape is obviously present, but weaker than the ones obtained for the magnetopause simulations.

Table caption: The measured electron and ion bulk speeds at the time reported in Figure 2 are compared with the $E\times B$ and diamagnetic speed for the same time interval. Not that the sampling rate of ions and electrons is different as is the time interval reported in Figure 2. For this reason the $E\times B$ speed of electrons and ions is not identical in this table.

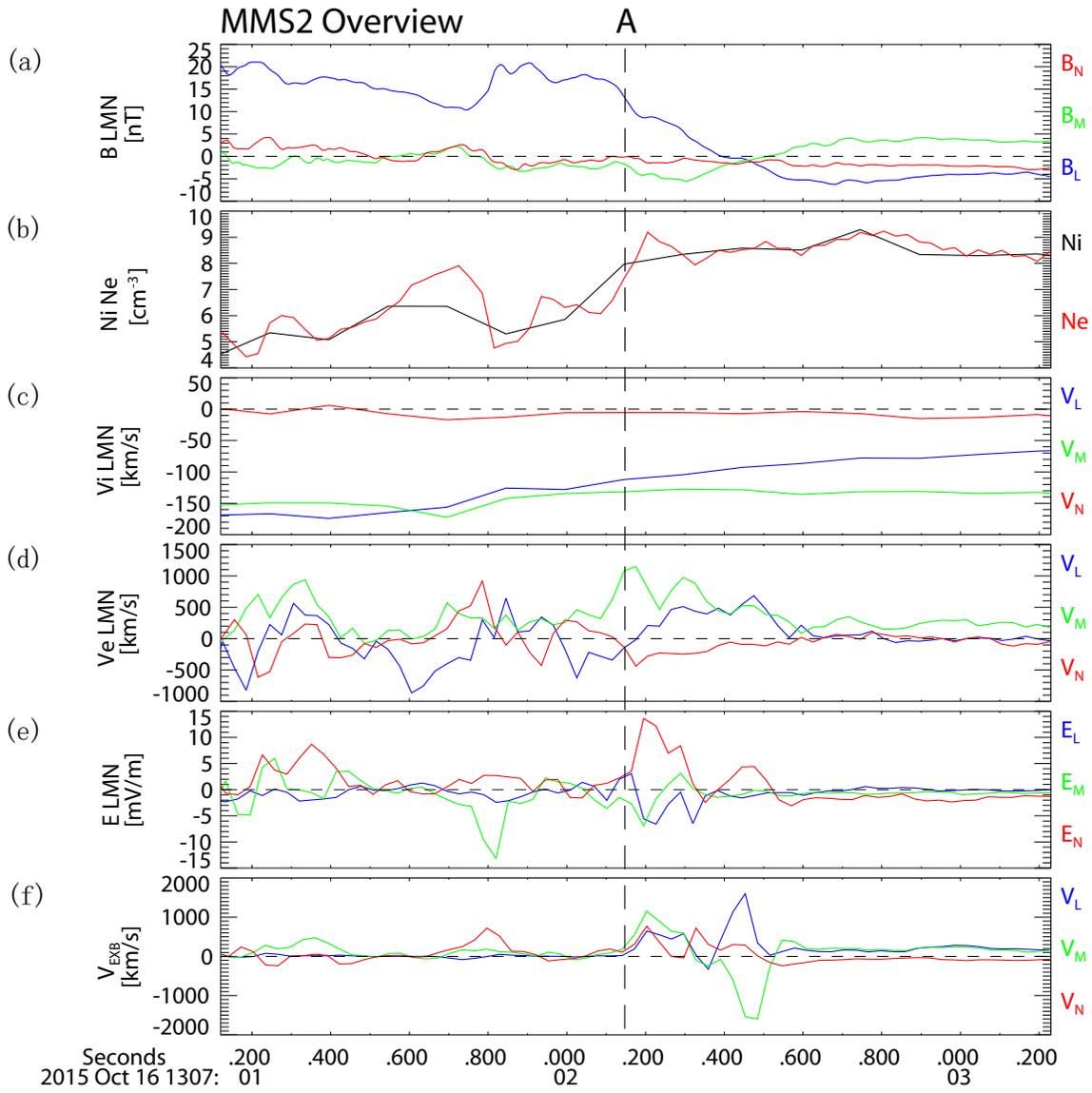

Figure 1: Magnetic field components (a), ion and electron density (b), ion velocity components (c), electron velocity components (d), electric field components (e) and $\boldsymbol{E} \times \boldsymbol{B}$ drift (f) from MMS2. All vectors are in the LMN boundary-normal coordinates defined by *Burch et al.* [2016]. N is the normal to the boundary and points away from the Earth, L corresponds to the main magnetic field component and is perpendicular to N and in the plane of reconnection (nearly along the magnetospheric magnetic-field direction), and M is normal to the L, N plane (predominantly dawnward). The time labeled as A is identified with the EDR crossing [*Burch et al.*, 2016].

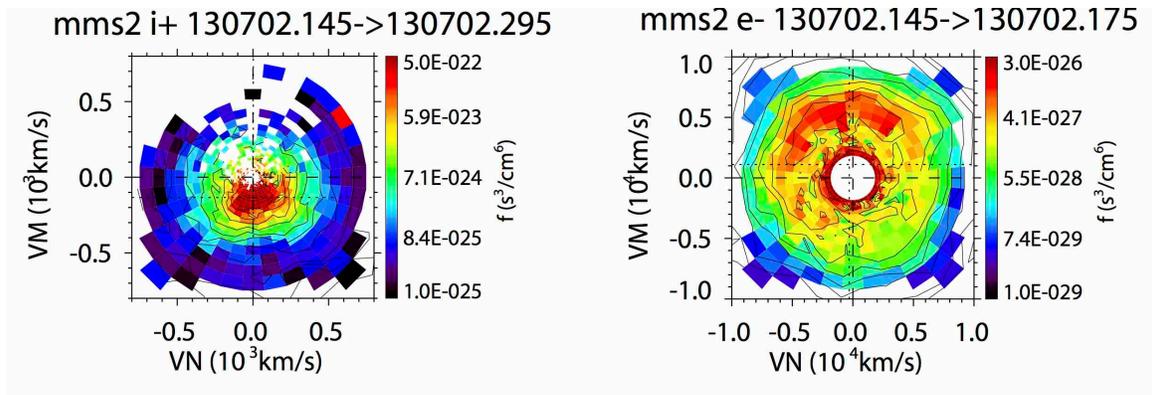

Figure 2: Electron and ion distribution functions at time A identified in Figure 1 displayed in magnetopause LMN boundary-normal coordinates defined by *Burch et al.* [2016]. The velocity plane $V_N$-$V_M$ is normal to the main magnetic field component $B_L$.

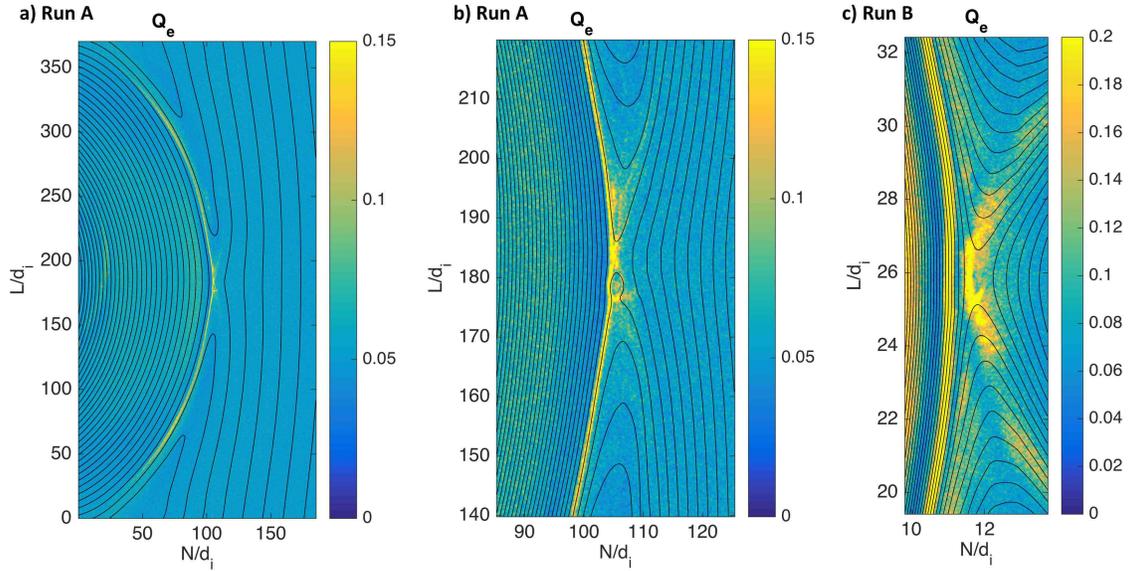

Figure 3: Characterization of the EDR from two multiscale simulations at the time $\omega_{ci,SP} t = 37$ (corresponding to the cyclotron frequency in the magnetosphere, or about 5 seconds of real time). Panels a and b are from Run A and panel c is from Run B (see the text for the parameters used in the runs). The non-gyrotropy values are computed by using the *Swisdak* [2016] definition. In all panels the black lines are magnetic surfaces computed from the out of plane vector potential. Run A shows the overall picture in the complete Run A domain (a) and in a zoom-in around the magnetopause subsolar region (panel b). Panel c shows the non-gyrotropy in a small subset of the Run B domain to show a close up of the EDR.

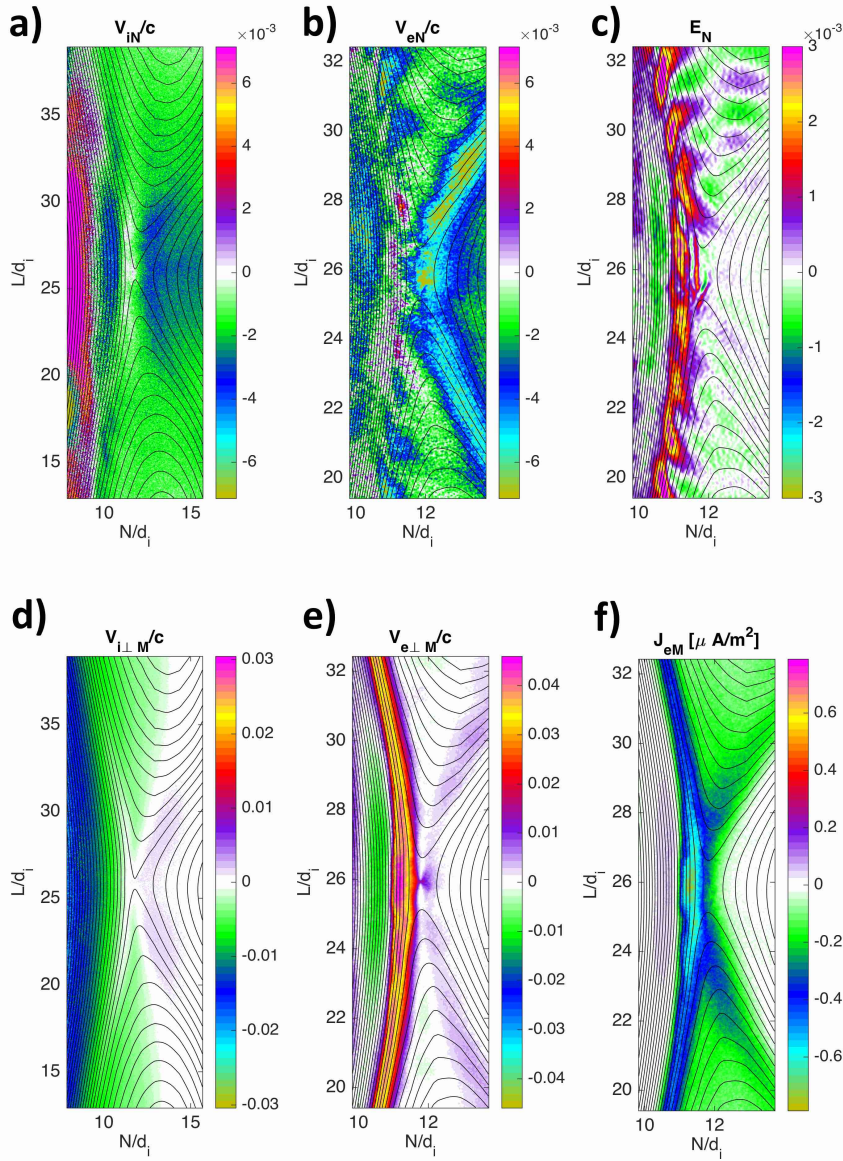

Figure 4: False color representation of the central region around the EDR observed in Run B. Panels a and b show the normal speed $V_{iN}$ and $V_{eN}$ for ions and electrons, respectively. Panel c shows the normal electric field $E_N$. The normal direction points sunward. Panels d and e show the out of plane perpendicular component of the velocity for ions and electrons, respectively. Panel f shows the out of plane (direction M) component of the electron current $J_{eM}$. The X-point is identified by the convergence of the magnetic surfaces shown in black while the stagnation point is identified by a zero value for the N-directed velocity: the stagnation point of electrons and ions is different; the ion panels (a, d) show a larger subdomain.

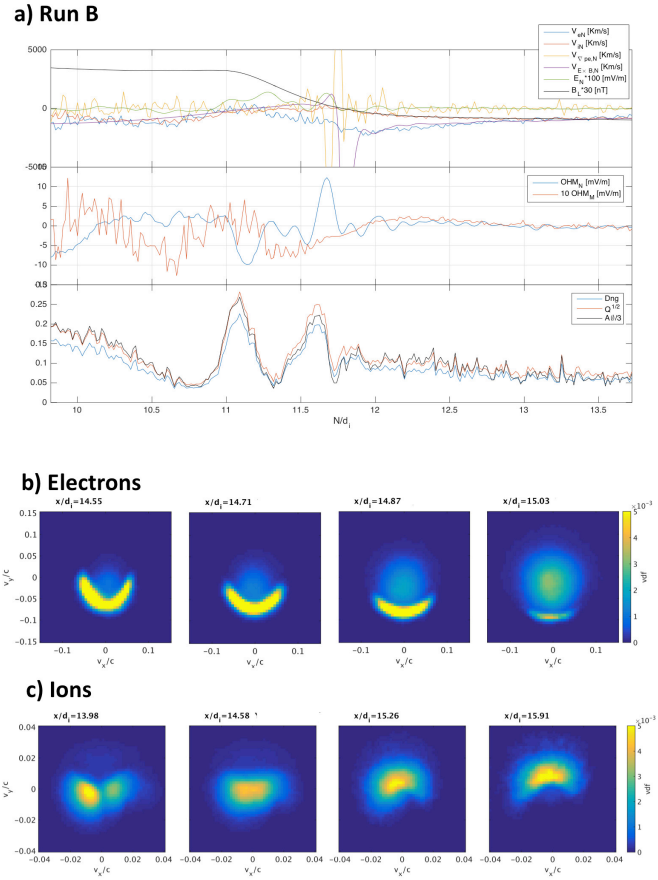

Figure 5: Panel a shows the traces of three physical quantities from run B which were taken along N in the vicinity of the EDR, and computed at the L position of the X-point (L/$d_i$ =25.68). On the top portion of panel a, the normal electric field, $E_N$, the normal velocity for electron (ions), $V_{eN}$ ($V_{iN}$), the normal $V_{ExB,N}$ drift, the electron diamagnetic drift $V_{de}$ and the magnetic field, $B_L$, are plotted. The middle part of panel a shows the violation of the electron frozen-in condition, $E + V_e \times B$ in the N and M directions. The bottom portion shows the three measures of non-gyrotropy discussed in the text (*Aunai et al.*, [2013] in blue, *Swisdak*, [2016] in red and *Scudder and Daughton* [2008] in orange). Panels b and c show the electron and ion velocity distributions in the perpendicular plane, respectively. The distributions are measured in a small box of size 0.15$d_{i,sh}$ around the position indicated above each distribution. The center of mass of the particles in the box is shown above each plot.

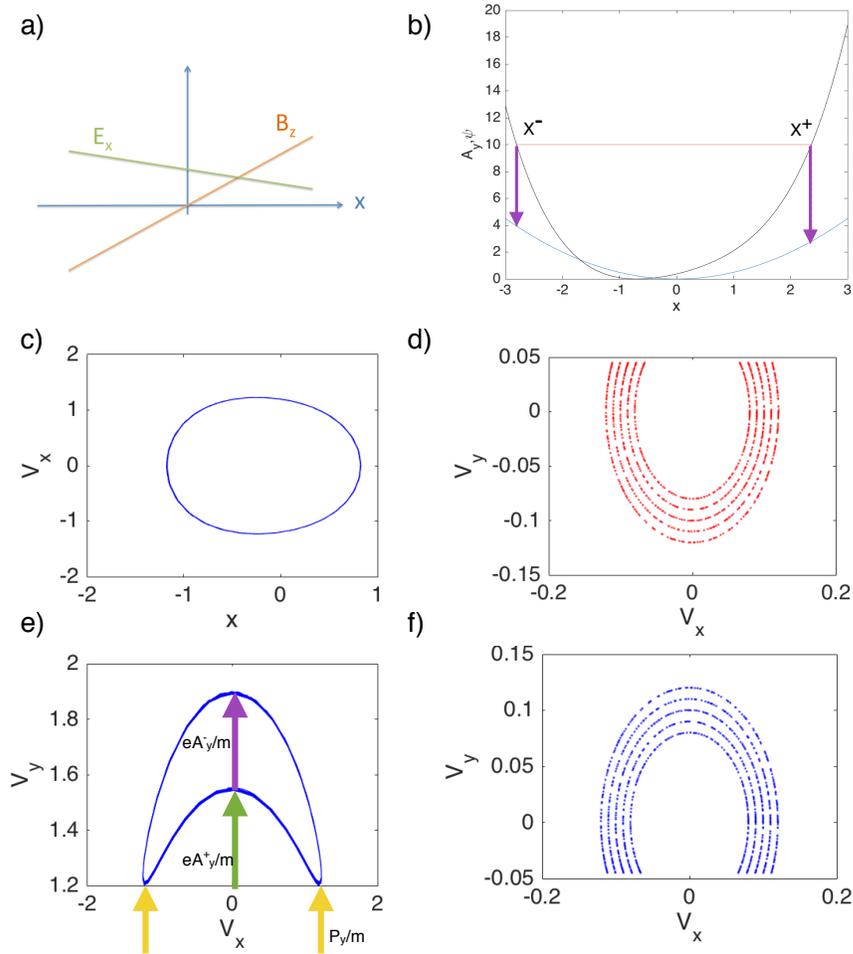

Figure 6: Adiabatic Hamiltonian model of the particle motion around the EDR. Panel a shows the idealized (linear) fields assumed in the model. Panel b shows one specific example of the pseudo-potential of the Hamiltonian model, the blue line is the vector potential $A_y$ and the black line is the pseudopotential $\psi$ derived from it. The model is not sensitive to the values, but just for presentation we chose specifically $B_0=0$, $B_1=1$, $E_0=-1$, $E_1=0$. A particle is confined by the well and oscillates between $X^-$ and $X^+$ where the vector potential $A_y$ takes the value indicated by the violet arrows. Panel c shows the x-$V_x$ phase space and panel e the perpendicular velocity plane $V_x$, $V_y$ for a particle. The particle forms a crescent shaped figure in the velocity plane, the value of the canonical momentum $p_y$ and of the vector potential at the two end points ($X^-$ and $X^+$) define the crescent. Panels d and f show a number of velocity plane trajectories of selected particles forming a crescent. The ions (red, panel d) and electrons (blue, panel f) are oppositely directed.

### a) L = EDR + 3.2 di

### b) L = EDR + 6.5 di

### c) L = EDR - 6.5 di

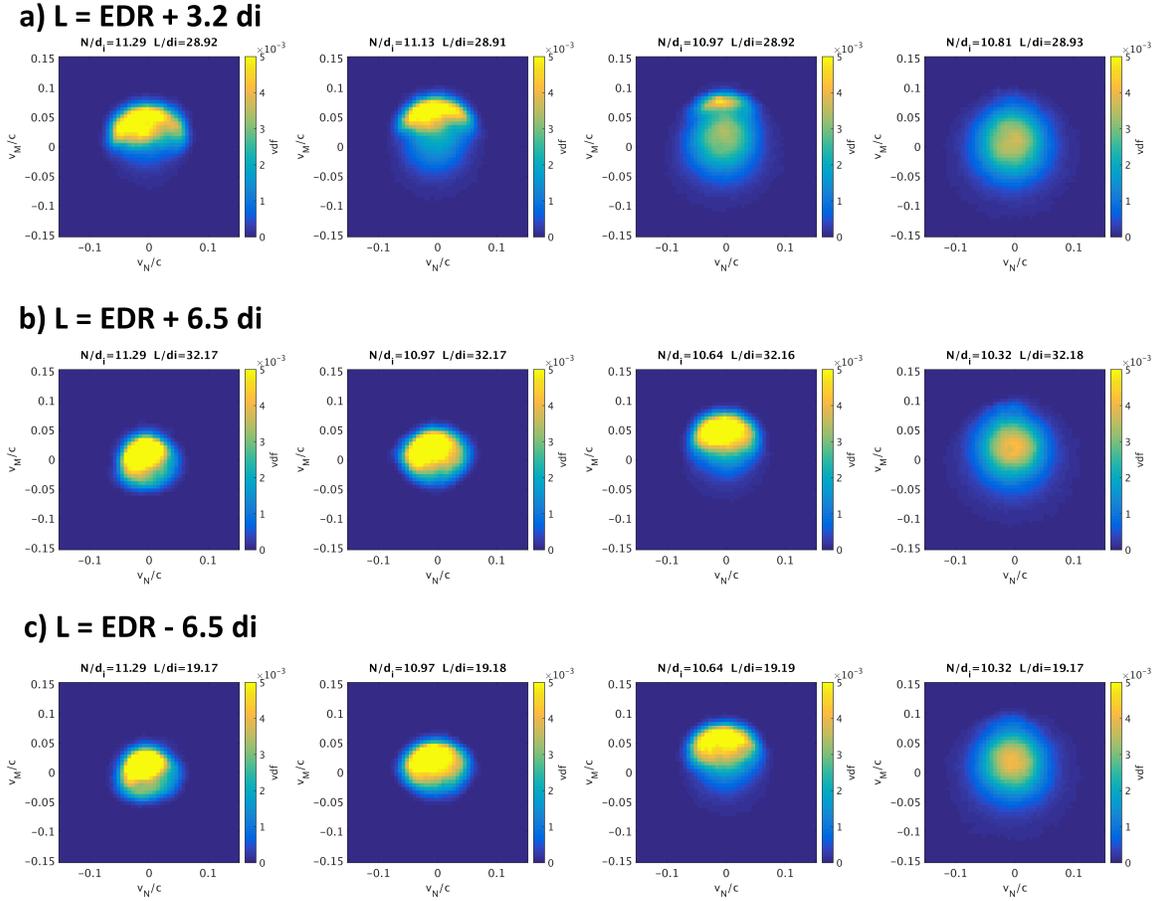

Figure 7: Electron velocity distribution in the perpendicular plane measured as in Figure 5b, in a small box of size $0.15 d_{i,sh}$ around the position indicated above each distribution. The distributions were determined at the same locations along the L-direction (indicated above each distribution) as in Figure 5 but at different north-south positions. Three sets of panels are shown. Panels in row a are taken around the $L/d_i = 28.92$, just north the EDR. Panels in row b and c are taken at two symmetric positions north and south of the EDR: $L/d_i = 32.17$ (north of the EDR) and $L/d_i = 19.17$ (south of the EDR).

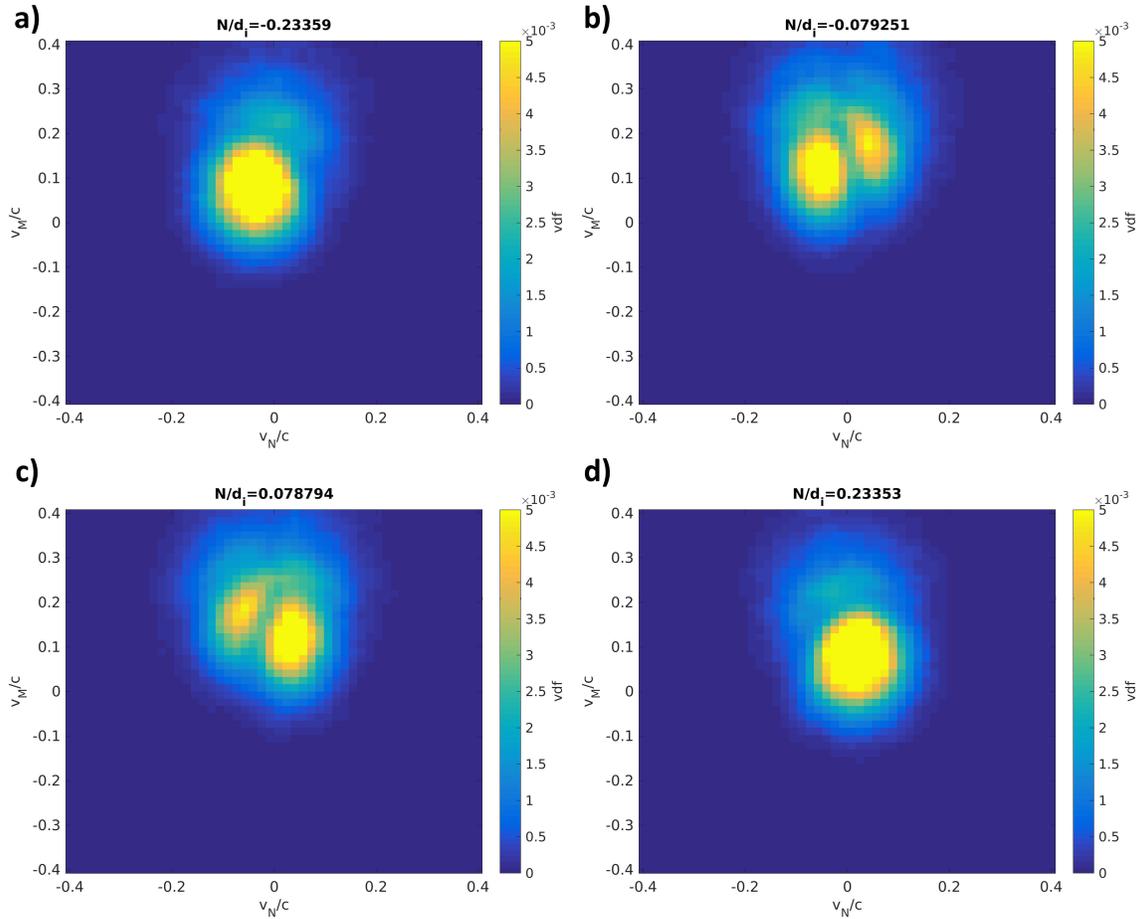

Figure 8: Electron distribution function in four different locations above and below the X-point (located at N=0) of a simulation for a symmetric reconnection geometry, which used a plasma with the same beta and temperature as that found in the magnetospheric side of Run B to form a Harris equilibrium (see text for detail). The plots are displayed in the perpendicular plane $V_M$-$V_N$. Velocity is normalized by the speed of light. The crescent shape is obviously present, but weaker than the ones obtained for the magnetopause simulations.